\documentclass{emulateapj}

\newcommand{\host}{SDSS~J0952+2143}
\usepackage{epstopdf}

\shorttitle{SDSS J0952+2143}
\shortauthors{Palaversa et al.}

\begin{document}

\title{Revealing the nature of extreme coronal-line emitter SDSS
J095209.56+214313.3}

\author{ Lovro Palaversa\altaffilmark{1},
Suvi Gezari\altaffilmark{2}, Branimir Sesar\altaffilmark{3},
J.~Scott Stuart\altaffilmark{4}, Przemyslaw
Wozniak\altaffilmark{5}, Berry Holl\altaffilmark{1},
\v{Z}eljko Ivezi\'{c}\altaffilmark{6} }

\altaffiltext{1}{Observatoire astronomique de l'Universit\'{e} de Gen\`{e}ve, 51
chemin des Maillettes, CH-1290 Sauverny, Switzerland\label{Geneve},
lovro.palaversa@unige.ch} 
\altaffiltext{2}{Department of Astronomy, University of Maryland, College Park,
MD 20742-2421, USA\label{UMD}}
\altaffiltext{3}{Max Planck Institute for Astronomy, K\"{o}nigstuhl 17, D-69117
Heidelberg, Germany\label{MPIA}}
\altaffiltext{4}{Lincoln Laboratory, Massachusetts Institute of Technology, 244
Wood Street, Lexington, MA 02420-9108, USA\label{LLMIT}} 
\altaffiltext{5}{Los Alamos National Laboratory, 30 Bikini Atoll Rd., Los
Alamos, NM 87545-0001, USA\label{LANL}} 
\altaffiltext{6}{University of Washington, Department of
Astronomy, P.O.~Box 351580, Seattle, WA 98195-1580, USA\label{Washington}}

\begin{abstract}

Extreme coronal-line emitter (ECLE) SDSSJ095209.56+214313.3, known by its
strong, fading, high ionization lines, has been a long standing candidate for a
tidal disruption event, however a supernova origin has not yet been ruled out.
Here we add several new pieces of information to the puzzle of the nature of the
transient that powered its variable coronal lines: 1) an optical light curve
from the Lincoln Near Earth Asteroid Research (LINEAR) survey that
serendipitously catches the optical flare, and 2) late-time observations of the
host galaxy with the Swift Ultraviolet and Optical Telescope (UVOT) and X-ray
telescope (XRT) and the ground-based Mercator telescope. The well-sampled,
$\sim10$-year long, unfiltered LINEAR light curve constrains the onset of the
flare to a precision of $\pm5$ days and enables us to place a lower limit on the
peak optical magnitude. Difference imaging allows us to estimate the location of
the flare in proximity of the host galaxy core. Comparison of the \textsl{GALEX}
data (early 2006) with the recently acquired Swift UVOT (June 2015) and Mercator
observations (April 2015) demonstrate a decrease in the UV flux over a $\sim 10$
year period, confirming that the flare was UV-bright. The long-lived UV-bright
emission, detected 1.8 rest-frame years after the start of the flare, strongly
disfavors a SN origin. These new data allow us to conclude that the flare was
indeed powered by the tidal disruption of a star by a supermassive black hole
and that TDEs are in fact capable of powering the enigmatic class of ECLEs.

\end{abstract} 


\section{Introduction\label{sec:intro}}

The class of extreme coronal line emitters (ECLEs) are distinct because they
exhibit strong coronal lines, such as [FeX] $\lambda$ 6376, [FeIX] $\lambda$
7894, [FeXIV] $\lambda$ 5304, that require a high-energy photoionizing
continuum \citep{kom08, Wang2011, Wang2012}.  Additionally, as a class, ECLEs
demonstrate coronal line intensities that show strong variability with time, as
well as complex Balmer-line profiles \citep{yan13}.  While tidal disruption
events (TDEs) have been proposed as the most likely source of the flaring
UV-soft X-ray photoionizing continuum powering the iron-line light echoes in
this class of objects, the light curve of the flare itself has never been
detected to test this scenario directly.

For the first time, we report the light curve of an ECLE, caught serendipitously
by the optical time-domain Lincoln Near Earth Asteroid Research \citep[LINEAR,
][]{sto00} survey, and provide Swift UV follow-up observations which confirm the
UV-luminous nature of the event.  The paper is organized as follows.  In \S 2 we
describe the new and archival observations of \host{}, in \S 3 we present the
implications for the TDE vs. supernova (SN) origin scenarios, and in \S 4 we
conclude that the coronal lines in \host{} were indeed the light echo of a TDE
in the gas-rich environment of a SMBH.

\section{Observations \label{sec:tech}}

Here we describe our new observations of \host{}, which together with the
extensive multi-wavelength data presented in \cite{kom09}, help solve the
mystery of its origin. Unless otherwise noted, when reporting observed
magnitudes we do not correct for Galactic extinction towards the source
(E(B-V)=0.028 mag). However, the absolute magnitudes include the correction for
Galactic extinction. Hereafter, we use UT dates, and assume a cosmology with
$H_0 = 70$ km s$^{-1}$ Mpc$^{-1}$, $\Omega_M = 0.3$, $\Omega_\Lambda = 0.7$, and
a luminosity distance of 360 Mpc.

\subsection{Archival data\label{sec:selection}}

\begin{figure*}
\epsscale{1}
\plotone{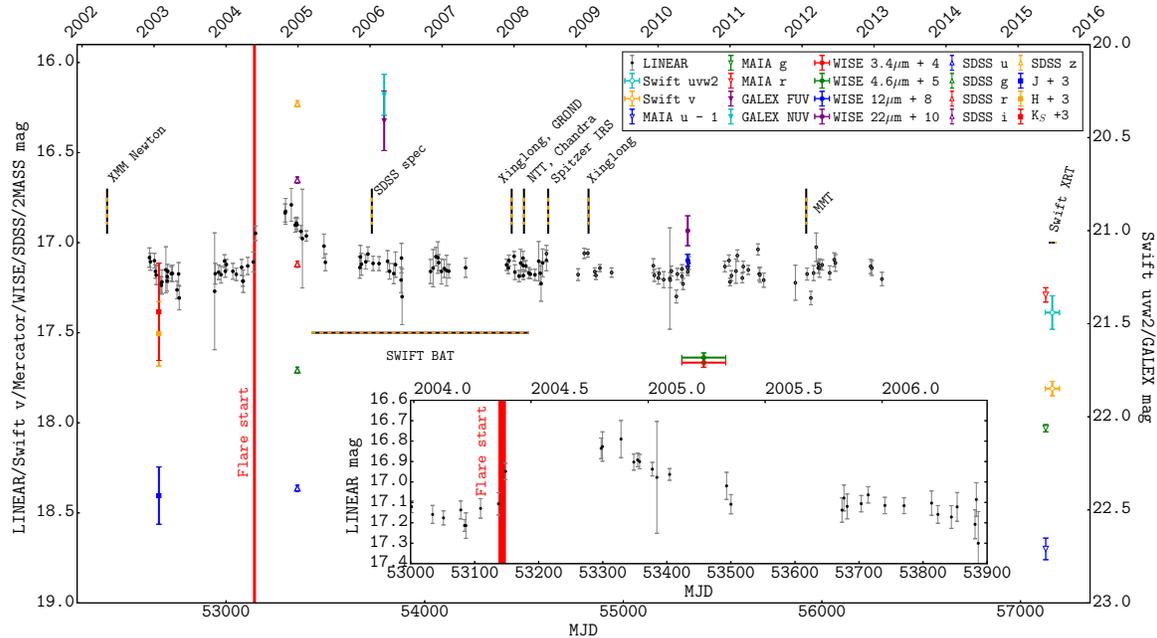}
\caption{Evolution of the flare and overplotted observations (symbols according
to the legend, note different scales on vertical axes). Time and duration of
spectroscopic observations, Swift BAT and Swift XRT are designated by black and
orange dashed lines. Red strip in the main panel and inset marks the limits on
the onset of the flare (2004 May 18 $\pm\ 5$ days). Note that the actual peak of
the optical light curve was not observed by LINEAR, and that earliest spectra
(SDSS) were obtained more than a year and a half after the peak of the flare.
Optical and UV magnitudes are expressed in AB system, 2MASS magnitudes
in the internal 2MASS system and WISE magnitudes in AB system. Galactic extinction was
not corrected for. A color version of this figure is available online.
\label{fig:LC}}
\notetoeditor{This figure needs to be stretched across two columns.}
\end{figure*}

During a systematic search for emission lines in AGNs in SDSS DR6 \citep[Sloan
Digital Sky Survey Data Release 6, ][]{sdssdr6}, \citet[][K08]{kom08} identified
unusual and variable emission lines of the host galaxy \host{} and subsequently
scheduled further observations with \textsl{Chandra X-ray Observatory} and the
Gamma-Ray Burst Optical/NIR Detector \citep[GROND,][]{gre08} instrument mounted
on the 2.2-m Max Planck Society telescope, as well as spectroscopic follow-up
with the {\it Spitzer Space Telescope} InfraRed Spectrograph
\citep[IRS,][]{hou04},  OMR spectrograph at the 2.16-m {\it Xinglong} telescope
and EMMI\footnote{see the EMMI user's manual at
\url{http://www.ls.eso.org/docs/}} instrument at the 3.5-m ESO New Technology
Telescope (NTT) \citep[see][K09]{kom09}. More recently, \cite[][Y13]{yan13}
acquired \host{} spectra from the Blue Channel Spectrograph on the Multi-Mirror
Telescope (MMT). The host \host{} galaxy was also observed by the ROSAT all-sky
survey in November 1990 \citep[RASS,][]{vog99}, 2MASS \citep{skr06},
\textsl{XMM-Newton} and \textsl{Swift} Burst Alert Telescope
\citep[BAT,][]{mar02, aje08} and \textsl{GALEX} \citep{mar05}. The chronology of
these observations is summarized in Fig.~\ref{fig:LC}.

The earliest ground-based spectrum, obtained by SDSS, exhibited unusually
strong, high ionization iron coronal lines with ionization states from [FeVII]
up to [FeXIV] and complex H$\alpha$ and H$\beta$ profiles that can be
decomposed into broad and narrow components with multiple peaks (c.f. K08, K09
and Y13 for a thorough discussion of spectral lines). Subsequent NTT, Xinglong
and MMT spectra revealed a dramatic fading of the these lines, as well as a
complex evolution of the broad-line profiles.

RASS (November 1990), \textsl{XMM-Newton} (May 7th 2002) and \textsl{Swift} BAT
(Mar 2005 -- Mar 2008) observations of the host galaxy did not make an X-ray
detection and were only able to place upper limits on the X-ray luminosity of
the host galaxy at $L_X(0.1-2.4~{\rm keV})<10^{43}$ erg s$^{-1}$,
$L_X(0.2-10~{\rm keV})<8\times10^{43}$ erg s$^{-1}$ and $L_X(15-55~{\rm
keV})<10^{44}$ erg s$^{-1}$, respectively. However, a {\it Chandra} 10 ks
observation initiated by K09 on 2008 Feb 4 detected faint X-ray emission, with
$L_X(0.1-10~{\rm keV}){}\sim{}10^{41}$ erg s$^{-1}$. The galaxy was detected as
a luminous MIR source by \textsl{Spitzer}, which K09 attribute to relatively
cold dust heated by the flare, or to a persistent starburst.

\subsection{LINEAR\label{sec:selection}}

The Lincoln Near Earth Asteroid Research \citep[LINEAR,][]{sto00} operated two
telescopes at the Experimental Test Site located within the US Army White Sands
Missile Range in central New Mexico at an altitude of 1506 m. The program used
two essentially identical equatorially mounted, folded design telescopes with
1-m diameter, f/2.5 primary mirrors equipped with 2560x1960 pixel
back-illuminated, frame transfer CCD cameras mounted in the prime focus. Cameras had no spectral filters and in combination with the telescopes produced a
$1\arcdeg.60\times1\arcdeg.23$ ($\approx2$ deg$^2$) field of view with a
resolution of $2.25\arcsec$/pix.

Spectral response curve of the LINEAR system peaks at approximately 625 nm and
covers approximately 400-1000 nm wavelength range, broadly matching the range of
SDSS $griz$ filters. \cite{ses11} described the LINEAR survey and photometric
recalibration based on SDSS stars acting as a dense grid of standard stars (for
the period from 1998--2009). In the overlapping ~10,000 deg$^2$ of sky between
LINEAR and SDSS, photometric errors range from ~0.03 mag for sources not limited
by photon statistics to ~0.20 mag at $r$ = 18 (where, $r$ is the SDSS $r$-band
magnitude). LINEAR photometry of the \host{} was obtained from
SkyDOT\footnote{\url{http://skydot.lanl.gov/}}.

In order to supplement the existing photometry (for the period after 2009) and
perform difference imaging, 622 $7.55\arcmin \times 7.55\arcmin$ (200 $\times$ 200 pixel)
image cutouts were extracted from the LINEAR database. Aperture photometry was
performed in the usual way using the IRAF\footnote{IRAF is distributed by the
National Optical Astronomy Observatories, which are operated by the Association
of Universities for Research in Astronomy, Inc., under cooperative agreement
with the National Science Foundation.} \citep{tod93} \texttt{apphot} task.
Images were astrometrically registered \citep[astrometry.net,][]{lan10} and then
visually inspected. Low quality frames in the non-flaring state were removed, to
give the final difference imaging sample of 299 images. Resulting good images
were divided into groups containing pre-, post- and flare data. Images
satisfying $53036>MJD>53750$ (2004 Jan 31st and 2006 Jan 14th) were then
corrected for distortions and co-added \citep[SWARP,][]{ber02} to create
template image from which the coadded image in the flaring state
($53148.2<MJD<53377.4$, 2004 May 22 to 2005 Jan 6) was differenced with HOTPANTS
\footnote{\url{http://www.astro.washington.edu/users/becker/v2.0\\/hotpants.html}},
an implementation of \cite{ala00} algorithm.

\subsection{Mercator Telescope \label{sec:merc}}

On 2015 Apr 15 we requested observations with the MAIA instrument mounted on the
1.2-m Mercator telescope\footnote{Located at Observatorio del Roque de los
Muchachos on La Palma island (Spain) and operated by Institute of Astronomy of
KU Leuven (Belgium)}. MAIA is an efficient three-channel imager, capable of
simultaneous three-band photometry. The optical system is built around three e2v
{2k$\times$6k} frame-transfer CCDs sourced from European Space Agency's canceled
Eddington mission. The field of view of the system is
9.4$\arcmin\times$14.1$\arcmin$, with image scale of 0.276$\arcsec$/pix. MAIA is
equipped with three filters: \textit{U, G} and \textit{R}. These filters are
similar, but not identical to SDSS filter system. In particular, the \textit{R}
filter is an approximation of SDSS $r+i$ filters, while \textit{U} and
\textit{G} are approximations of SDSS $u$ and $g$. More details can be found in
technical paper by \cite{ras13}.

\textit{U, G} and \textit{R} band images were acquired simultaneously and under
good conditions on 2015 Apr 15th, with a 1 ks exposure in all three filters.
Usual reduction steps were performed by a custom built Python script (Palaversa
\& Blanco-Cuaresma 2015, private comm.) that is used for reduction of Gaia
Science Alerts\footnote{\url{http://gaia.ac.uk/selected-gaia-science-alerts}}
(GSA) follow-up observations obtained by the MAIA instrument. The photometric
calibration and conversion to the SDSS system were performed by the Cambridge
Photometry Calibration Server (also a part of GSA). The host galaxy was
detected with $u=19.70\pm0.06$ mag, $g=18.03\pm0.02$ mag and $r=17.29\pm0.04$
mag. Comparison with SDSS photometry of 2004 Dec 20 ($u=18.36\pm0.02$ mag,
$g=17.71\pm0.01$ mag, $r=17.119\pm0.005$ mag, $i=16.652\pm0.005$ mag and
$z=16.23\pm0.01$ mag) reveals that the source became fainter in the $u$, $g$,
and $r$ bands, and redder (${(u-g)}_{SDSS}=0.66$, ${(g-r)}_{SDSS}=0.59$,
${(u-g)}_{MAIA}=1.67$, ${(g-r)}_{MAIA}=0.74$). All magnitudes are in AB
system.
\newpage
\subsection{Swift\label{sec:selection}}
\begin{figure}
\epsscale{1}
\plotone{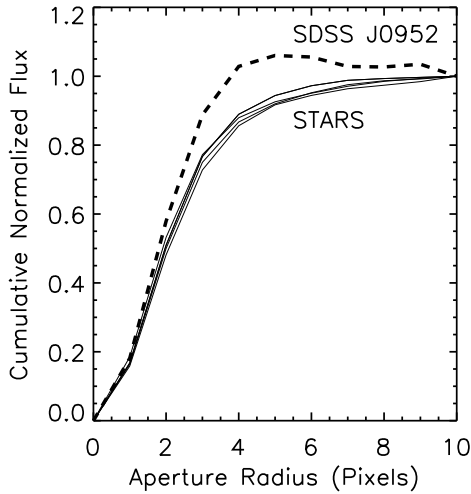}
\caption{Cumulative flux as a function of aperture radius for SDSS J0952+2143 
(dashed line) in comparison to reference stars (solid lines) in the Swift 
$uvw2$ image.  The UV emission in SDSS J0952+2143 is clearly extended, and is 
likely associated with star-formation in the host galaxy. The Swift
UVOT pixel scale is 0.502\arcsec/pix. \label{fig:rad}}
\end{figure}

We obtained follow-up imaging of \host{} with Swift UVOT during the time period
of 2015 April 12 -- June 23 with 4.96 ksec in the $uvw2$ ($\lambda_{\rm eff} =
2246$ \AA) filter and 4.79 ksec in the $v$ ($\lambda_{\rm eff} = 5468$ \AA)
filter. The host galaxy is detected with $uvw2 = 21.44 \pm 0.09$ mag and $v =
17.81 \pm 0.04$ in the AB system using heasoft software package {\tt uvotsource}
and a $5\farcs0$ radius aperture. We find a negligible correction ($< 0.1$ mag)
between AB magnitudes in the \textsl{GALEX} $NUV$ and Swift $uvw2$ bands using a
comparison of four reference stars detected in both the Swift image and the 218
sec \textsl{GALEX} All-sky Imaging Survey (AIS) obtained on 2006 March 2.
\host{} is spatially extended in the $uvw2$ image in comparison to reference
stars in the field of view (see cumulative flux distribution in Figure
\ref{fig:rad}).  We also measure a 3$\sigma$ upper limit on the X-ray flux from
a 6.82 ksec Swift XRT exposure on 2015 June 23 using the heasoft software
package {\tt sosta} of $f_{\rm 0.3-10 keV} < 9.54 \times 10^{-14}$ ergs
s$^{-1}$, which for a Galactic column density of $N_{H} = 2.79 \times 10^{20}$
cm$^{-2}$ and a power-law index of $\Gamma = 1.9$, translates to $L_{\rm X} <
1.47 \times 10^{42}$ ergs s$^{-1}$.  This upper limit is consistent with the
much more sensitive late-time \textsl{Chandra} detection of \host{} on 2008
February 4.

\subsection{WISE\label{sec:WISE}}

Wide-field Infrared Survey Explorer \citep[WISE,][]{2010AJ....140.1868W} is a
0.4 m NASA infrared wavelength space telescope in Earth orbit that performed an
all-sky survey in $3.4\ \mu$m, $4.6\ \mu$m, $12\ \mu$m and $22\ \mu$m (hereafter
designated as $W1$, $W2$, $W3$ and $W4$, respectively). WISE detected the host
galaxy with $W1=13.67\pm0.03$ mag, $W2=12.64\pm0.03$ mag, $W3=9.10\pm0.03$ mag
and $W4=6.93\pm0.08$ mag, between 2010~May~08 and 2010~Nov~15. All
magnitudes are in Vega system.

\section{Analysis \label{sec:analysis}}

Ten-years long monitoring enabled by the LINEAR survey as well as late-time
Swift and MAIA observations allow us to uncover critical diagnostic information missing
from the previous analyses of the mechanism responsible for the luminous flare
in the SDSS J095209.56+214313.3 galaxy.

Most importantly, exact timing of the event and its evolution can now be
constrained (see Fig.~\ref{fig:LC}). From the difference in time between the
last point on the flat part of the light curve prior to the flare and the first point
on the rise, we estimate with a precision of $\pm$5 days that the flare started
on 2004 May 18. Unfortunately LINEAR did not observe the peak of the optical
light curve, but we are able to determine that the flare could not have been
fainter than $M_r\sim-20$ mag.

\begin{figure}
\epsscale{1}
\plotone{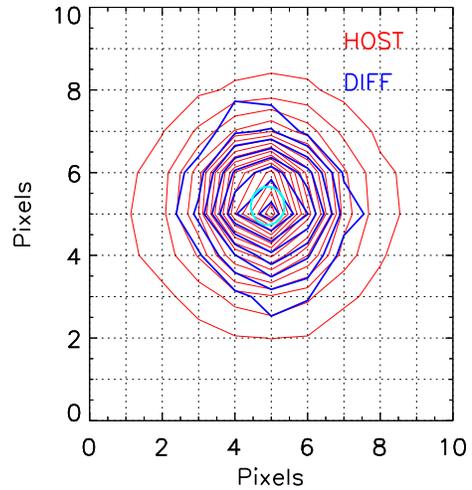}
\caption{Contour image of the host galaxy detected by LINEAR constructed from
the pre- and post-flare data (red), compared to the contour image of the
difference image during the flare (blue).  Cyan circle marks the 2$\sigma$ error
circle on the difference image centroid measured from a Gaussian fit. No
significant offset is detected between the difference image centroid and the
host galaxy nucleus. The LINEAR pixel scale is 2.25\arcsec/pix. A color
version of this figure is available online. \label{fig:offset} \\}
\end{figure}

LINEAR survey also allows us to establish the optical variability of the host
galaxy \host{} at a level of $\sigma<0.08$ mag (outside of the flaring phase),
removing the possibility of strong, unobscured AGN activity in the host. We also
use the difference imaging described in \S\ref{sec:selection} to localize the
transient relative to the host galaxy nucleus.  Figure~\ref{fig:offset} shows a
contour of the difference image constructed from the flaring state images,
overlaid on the host galaxy reference image.  There is no significant offset
detected, with the transient centroid measured from a Gaussian fit located
within 1$\sigma$ (0.2 pixels or 0.45$\arcsec$) of the host galaxy centroid. 
Assuming the SDSS value for redshift ($z$=0.079), this translates to an offset
from the core of less than 670 pc, thus not ruling out the TDE hypothesis.

Furthermore, the optical light curve allows us to put other observations in context.
We now know that the SDSS photometry (2004 Dec 20) and SDSS spectrum (2005 Dec
30) were taken approximately 210 and 580 days after the start of the flare.
Therefore, only LINEAR and SDSS photometry were taken during the flaring phase.
By the time \textsl{GALEX} photometry and SDSS spectra were acquired, the flare had
already faded considerably (at least at optical wavelengths). Although the early
part of the Swift BAT observations were taken during the declining phase of the
flare, no detection was made suggesting that there was no significant hard X-ray
emission. Remaining observations taken after mid-2006 could have measured only
the echo of the flare in the surrounding medium.

The extended, persistent UV emission detected by Swift from \host{} $11$ years
after the transient outburst, has faded by 1.2 mag since the \textsl{GALEX}
detection on UT 2006 March 02, and is likely associated with star formation in
the host galaxy.  Therefore, we can derive the UV flux intrinsic to the
transient detected by \textsl{GALEX} to be $NUV_{\rm trans} = 20.60 \pm 0.14$
mag, which corresponds to an absolute magnitude of $M_{NUV} = -17.4$~mag
(corrected for Galactic extinction), 1.8 years after the onset of the
transient outburst. Our confirmation of transient UV emission associated with the event in
2006 March 2 contradicts the conclusions of \cite{yan13}, who found that it is
not necessary to add a non-stellar component to fit the blue end of the SDSS
spectrum of \host{} on 2005 December 30, which we now know was taken 1.6 yr
after the start of the UV/optical flare.  Note that a similar fading in the UV
was detected in archival \textsl{GALEX} observations of ECLE SDSS J0748+4712 by
\citet{Wang2012}, confirming that it too was powered by a UV-luminous event.

Furthermore, the confirmation that the host galaxy is in fact a star-forming
galaxy is important, since the continuum colors of the host galaxy measured by
2MASS, WISE and MAIA, all of which were taken either before the flare, or at
least six years after the optical light curve reached its pre-flare level, and
its narrow-line ratios measured in K08, are ambiguous, and consistent with the
regions populated by both AGN and star-forming galaxies in the diagnostic
diagrams \citep{obr06, nik14, bpt81}.

\section{Discussion}

\subsection{Nature of the Host Galaxy}

LINEAR photometry spanning more than a decade does not exhibit behavior
indicative of AGN activity (c.f. stability of the optical light curve in
\S{\ref{SEC:ANALYSIS}}). This is corroborated by \textit{XMM Newton, Swift
BAT, Swift XRT} and \textit{Chandra} X-ray observations, all of which were taken
outside of the flaring phase (with the slight exception of Swift BAT
observations). Only \textit{Chandra} observations detected low levels of soft
X-rays, approximately 3.5 years after the start of the flare, and at a level
well below that which is expected for normal AGN. Our recent \textsl{Swift}
photometry detected extended, persistent UV emission from \host, an indication
of ongoing star formation. Given the fact that the Swift photometry was acquired
$\sim$11 years after the start of the flare, it is unlikely that there is a
contribution to the late-time UV emission from the flare itself.

\begin{figure}
\epsscale{1.18}
\plotone{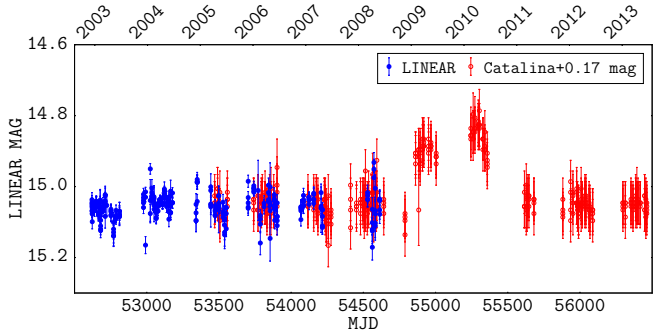}
\caption{IC3599 optical light curve. Blue empty circles correspond to the
LINEAR data and red empty circles to CSDR2 data \citep{dra09}, shifted by
0.17~mag, i.e. the difference of the median LC values between LINEAR and CSDR2
data, outside the flaring phase. Please note that \host~is $\sim$2~mag fainter
than IC3599 and actually near the faint limit of the LINEAR survey.
\label{fig:ic3599}}
\notetoeditor{This figure needs to be stretched across two columns.}
\end{figure}

These measurements provide a contigous observational baseline of
approximately two years before the flare and ten years after the flare in which
no activity characteristic of AGN was detected. We also note that RASS
observations in Nov 1990 did not detect significant X-ray emission (yet an
11-year gap in observation exists between RASS and XMM-Newton observations).
However, a longer baseline may be needed to definitively rule out an AGN.
Seyfert 1.9 galaxy IC 3599, for example, showed two bursting episodes caught by
ROSAT and Swift, respectively, separated by a time interval of 20 years. While
the Catalina Sky Survey data caught the second outburst in 2008 \citep{gru15},
it shows no significant variability ($\sigma = 0.04$ mag) in the LINEAR data in
the 6 years preceding it (Fig.~\ref{fig:ic3599}). We do, however, note the
difference in the LC shape of IC 3599 and \host, the latter of which is
asymmetric, shows more abrupt change in luminosity and has a larger
optical amplitude (approximately 0.2 mag in case of IC 3599 and 0.5 mag in
case of \host).

Thanks to the LINEAR light curve, [NII]/H$\alpha$ and [OIII]/H$\beta$ line
ratios can now be used more safely in the context of BPT diagram, in which the
host galaxy is placed in the SF region, but near the AGN/SF division. Similarly,
2MASS, MAIA, SDSS and WISE color-color diagrams locate the host galaxy within
the clusters occupied by AGN and SF galaxies. Motivated by these clues (and by
GROND imaging that shows spiral structure in the host (K09)), we conclude that
the flare probably happened in a non-active, star forming galaxy.

\subsection{Nature of the Flare}\label{sec:flare}

\begin{figure}
\hspace{-1.cm}
\epsscale{1.25}
\plotone{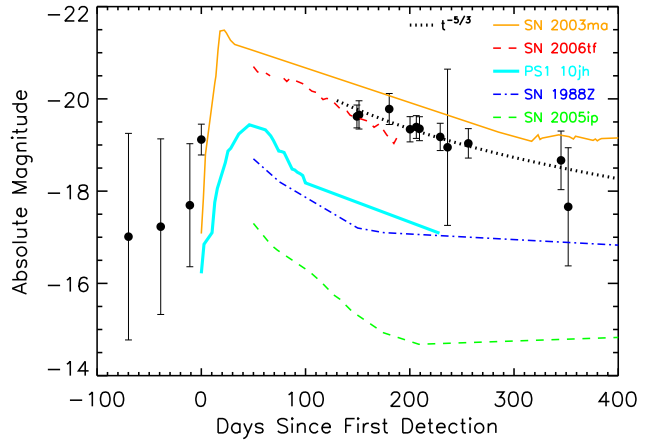}
\caption{Comparison of LINEAR difference imaging light curve, converted to
absolute magnitudes, to known extreme interacting SNe SN 2005ip (unfiltered), SN
1988Z ($M_z$), SN 2006f ($M_r$), SN 2003ma ($M_r$) and the prototypical TDE
candidate PS1-10jh ($M_r$). Correction for Galactic extinction is included. A
color version of this figure is available online.
\label{fig:sne}}
\end{figure}

There are several possible explanations for the outburst in \host{}: a
tidal disruption of a star by a supermassive black hole in the center of the
host galaxy, an extreme Type IIn SN or AGN-like variability. Similarities in
the spectral line responses of the former two scenarios require that we look at
the photoionizing flare itself to ultimately uncover its origin. In Figure
\ref{fig:sne}, we compare the LINEAR difference imaging light curve converted to
absolute magnitude, with the light curves of extreme interacting SNe in the
$r$-band (SN 2003ma, \cite{Rest2011}; SN 2006tf, \cite{smi08}; SN 1988Z,
\cite{Turatto1993}; SN 2005ip, \cite{smi09}), and the best-observed TDE
candidate PS1-10jh \citep{gez12}.  While the decline rate in the optical of 0.57
mag/(100 days) at $> 150$ days from the start of the flare is similar to the
behavior of SNe 2003ma and 2006tf, this decline can also be fitted with a
$t^{-5/3}$ power-law evident in the optical light curve of PS1-10jh.  MAIA
photometry obtained $\sim$10 years after SDSS photometry shows that largest
change in brightness happened in the bluer bands ($\Delta u \sim$1.3 mag and
$\Delta g \sim$ 0.6 mag), indicating that the flare itself was much bluer than
the host galaxy.

The coronal line formation, however, was a response to intense X-ray radiation
created by the event. Since {\it Chandra} spectrum was taken $\sim$ 3.5 years
after the peak of the flare it is reasonable to expect that the X-ray luminosity
of \host{} could have been orders of magnitude larger (assuming the
$t^{-5/3}$ decay predicted for TDEs). Such high levels of X-ray luminosity
are unusual for SNe. The inferred UV luminosity for \host{} at 1.8 yr after the
start of the flare is comparable to the late-time UV luminosity inferred for TDE
candidate PS1-10jh at a similar phase \citep{Gezari2015b}, and much more
luminous than would be expected for an interacting SN at such late times.  UV
observations of SNe show a dramatic fading in the $NUV$ on the timescale of days
to weeks \citep{Brown2009, Gezari2015a}, due to the expansion and cooling of the
SN ejecta.  Even in interacting SNe, sustained $NUV$ emission on the level
observed for \host{} on the timescale of 1.8 yr after explosion would be
unprecedented.

A possible explanation for the outburst in the \host~could be AGN
variability, such as an accretion disk instability \citep[e.g.][]{gru15}. Thanks
to the LINEAR and Catalina surveys, we are able to directly compare the optical
light curve of SDSS J0952+2143 to the light curve of the bursting AGN IC 3599.
This comparison indicates a shorter rise times in SDSS J925+2143, more
consistent with the TDE scenario than a disk instability (Saxton et al. 2015)
and a decline with the $t^{-5/3}$ rate (unlike the case of IC 3599).
Furthermore, while the ionization lines up to [FeX] have been reported in case of IC 3599,
the extreme coronal lines with ionization up to [FeXIV] were not detected.  We
find that the reported X-ray and UV flux levels, combined with the light curves,
suggest the AGN scenario to be less likely than a TDE for SDSS J0952+2143.
However, a longer observational baseline of optical monitoring (to be provided
by e.g. Pan-STARRS, Gaia, and LSST) could shed more light on this question.

Interestingly, the peak optical luminosity of the transient is $>$ 1 mag
brighter than PS1-10jh \citep{gez12}, but in the range of some other
optically-selected TDE candidates reported in the literature
\citep{vanVelzen2011, Arcavi2014}.  This could be attributed to a more energetic
event, possibly resulting from the efficient accretion of a larger fraction of
the bound stellar debris.  This larger luminosity could also translate to a
stronger light echo in its surrounding gaseous environment, thus resulting in a
detectable ECLE.  Given the optical luminosity of the host galaxy measured by
\textsl{Swift} and the Mercator telescope and its stellar velocity dispersion,
a central black hole with a mass of $7\times10^6 M_\odot$ \citep{kom08}
is well within the range of black hole masses capable of disruption of a
solar-type star outside the event horizon, and in a mass range where the peak
accretion rate of the stellar debris would not be limited by the Eddington
luminosity of the central SMBH.

\section{Conclusions}

We present archival LINEAR observations, and late-time Swift and Mercator
observations, that add new pieces of the puzzle to the nature of the flare that
powered the extreme coronal line emission in \host{}. In particular, our
observations reveal the blue color and strong late-time UV luminosity of the
flare, and constrain its location within the errors to the nucleus of the host
galaxy, both disfavoring a SN origin.  Furthermore, if the flaring event is
indeed associated with the galaxy's central SMBH, then the lack of variability
detected by LINEAR before and after the flare is best explained by an impulsive
accretion event, as would be expected from the tidal disruption of a star, as
opposed to stochastic variability associated with a persistently accreting AGN.
In Table 1 we summarize the evidence for the nature of the flaring event that
powered the light echo in \host{}, and conclude that the most likely scenario
that explains all of its properties is a TDE.

\begin{deluxetable}{lccc}
\centering
\tabletypesize{\scriptsize}
\tablecaption{Compatibility of the observations with two competing scenarios
(a TDE, an interacting supernova or an AGN). ``Y'' indicates ``yes'', ``N''
indicates ``no'' and ``U'' indicates ``unlikely''.\label{tab:pro_con} }
\tablewidth{210pt}
\tablehead{
\colhead{Property}  & \colhead{TDE} & \colhead{SN}  & \colhead{AGN}}
\startdata
Proximity to host core		& Y		    &  U		    &  Y \\
UV luminosity				& Y	    	&  N 		    &  Y \\
X-ray luminosity			& Y		    &  U		    &  N \\
LC shape					& Y		    &  Y		    &  U
\enddata
\notetoeditor{Please move the table to Section 5: Conclusions, and add some
vertical space underneath.}
\end{deluxetable}

This case of \host{} demonstrates the importance of archived all-sky,
time-domain surveys: LINEAR was originally an asteroid survey that was recycled
as a project searching for variable stars. However, this resulted in a highly
valuable archival, decade long, time-domain survey covering a large fraction of
the sky. In the future era of synoptic surveys, the recovery of the light curves
of ECLEs discovered in spectroscopic surveys should be even easier, and allow
one to relate the detailed energetics of the TDE powering the flare, to its
subsequent light echo in the gaseous environment of the SMBH.

\section{Acknowledgments\label{sec:ack}}

L.P. acknowledges support by the Gaia Research for European Astronomy Training
(GREAT-ITN) Marie Curie network, funded through the European Union Seventh
Framework Programme ([FP7/2007-2013] under grant agreement n$\degr$ 264895 and
valuable conversations with Laurent Eyer. S.G. was supported in part by NASA
Swift grant NNX15AR46G, and by NSF CAREER grant 1454816. Authors would also like
to thank the anonymous referee for the useful comments.

The LINEAR program at MIT Lincoln Laboratory is funded by the National
Aeronautics and Space Administration Near-Earth Object Observations Program via
an interagency agreement under Air Force Contract \#FA8721-05-C-0002.  Opinions,
interpretations, conclusions and recommendations are those of the authors and
are not necessarily endorsed by the United States Government. Mercator Telescope
is operated on the island of La Palma by the Flemish Community, at the Spanish
Observatorio del Roque de los Muchachos of the Instituto de Astrof\'{i}sica de
Canarias. MAIA camera was built by the Institute of Astronomy of KU Leuven,
Belgium, thanks to funding from the European Research Council under the European
Community's Seventh Framework Programme (FP7/2007-2013)/ERC grant agreement no
227224 (PROSPERITY, PI: Conny Aerts) and from the Fund for Scientific Research
of Flanders (FWO) grant agreement G.0410.09. The CCDs of MAIA were developed by
e2v in the framework of the ESA Eddington space mission project; they were
offered by ESA on permanent loan to KU Leuven. Photometric calibrations for MAIA
instrument were obtained using the Cambridge Photometric Calibration Server
(CPCS), designed and maintained by Sergey Koposov and \L ukasz Wyrzykowski.
This publication makes use of data products from the Wide-field Infrared Survey
Explorer, which is a joint project of the University of California, Los Angeles,
and the Jet Propulsion Laboratory/California Institute of Technology, funded by
the National Aeronautics and Space Administration. Funding for the SDSS and
SDSS-II has been provided by the Alfred P. Sloan Foundation, the Participating
Institutions, the National Science Foundation, the U.S. Department of Energy,
the National Aeronautics and Space Administration, the Japanese Monbukagakusho,
the Max Planck Society, and the Higher Education Funding Council for England.
The SDSS Web Site is http://www.sdss.org/. The SDSS is managed by the
Astrophysical Research Consortium for the Participating Institutions. The
Participating Institutions are the American Museum of Natural History,
Astrophysical Institute Potsdam, University of Basel, University of Cambridge,
Case Western Reserve University, University of Chicago, Drexel University,
Fermilab, the Institute for Advanced Study, the Japan Participation Group, Johns
Hopkins University, the Joint Institute for Nuclear Astrophysics, the Kavli
Institute for Particle Astrophysics and Cosmology, the Korean Scientist Group,
the Chinese Academy of Sciences (LAMOST), Los Alamos National Laboratory, the
Max-Planck-Institute for Astronomy (MPIA), the Max-Planck-Institute for
Astrophysics (MPA), New Mexico State University, Ohio State University,
University of Pittsburgh, University of Portsmouth, Princeton University, the
United States Naval Observatory, and the University of Washington. The CSS
survey is funded by the National Aeronautics and Space Administration under
Grant No. NNG05GF22G issued through the Science Mission Directorate Near-Earth
Objects Observations Program.  The CRTS survey is supported by the U.S.~National
Science Foundation under grants AST-0909182.


\begin{thebibliography}

\bibitem[Adelman-McCarthy et al.(2008)]{sdssdr6} Adelman-McCarthy, J.~K., Ag{\"u}eros, M.~A., Allam, S.~S., et al.\ 2008, \apjs, 175, 297
\bibitem[Ajello et al.(2008)]{aje08} Ajello, M., Greiner, J.,
Kanbach, G., et al.\ 2008, \apj, 678, 102 
\bibitem[Alard(2000)]{ala00} Alard, C.\ 2000, \aaps, 144, 363
\bibitem[Arcavi et al.(2014)]{Arcavi2014} Arcavi, I., et al.~2014, \apj, 793, 38
\bibitem[Aretxaga et al.(1999)]{are99} Aretxaga, I., Benetti, S.,
Terlevich, R.~J., et al.\ 1999, \mnras, 309, 343
\bibitem[Bade et al.(1996)]{bad96} Bade, N., Komossa, S., \& Dahlem, M.\ 1996,
\aap, 309, L35 
\bibitem[Baldwin et al.(1981)]{bpt81} Baldwin, J.~A., Phillips,
M.~M., \& Terlevich, R.\ 1981, \pasp, 93, 5
\bibitem[Bertin et al.(2002)]{ber02} Bertin, E., Mellier, Y., Radovich, M., et al.\ 2002, Astronomical Data Analysis Software and Systems XI, 281, 228
\bibitem[Breeveld et al.(2011)]{Breeveld2011} Breeveld, A.~A., Landsman, W., Holland, S.~T., Roming, P., Kuin, N.~P.~M., \& Page, M. J. 2011, AIPC, 1358, 373
\bibitem[Brown et al.(2009)]{Brown2009} Brown, P.~J., et al.~2009, \aj, 137, 4517
\bibitem[Burrows et al.(2011)]{bur11} Burrows, D.~N., Kennea, J.~A., Ghisellini,
G., et al.\ 2011, \nat, 476, 421 
\bibitem[Chandra et al.(2012)]{cha12} Chandra, P., Chevalier,
R.~A., Chugai, N., et al.\ 2012, \apj, 755, 110
\bibitem[Drake et al.(2009)]{dra09} Drake, A.J. et al. First Results from the
Catalina Real-time Transient Survey 2009, ApJ, 696, 870
\bibitem[Esquej et al.(2007)]{esq07} Esquej, P., Saxton, R.~D.,
Freyberg, M.~J., et al.\ 2007, \aap, 462, L49 
\bibitem[Fabian \& Terlevich(1996)]{fab96} Fabian, A.~C., \&
Terlevich, R.\ 1996, \mnras, 280, L5 
\bibitem[Fox et al.(2000)]{fox00} Fox, D.~W., Lewin, W.~H.~G.,
Fabian, A., et al.\ 2000, \mnras, 319, 1154
\bibitem[Gezari et al.(2009)]{gez09} Gezari, S., Halpern, J.~P.,
Grupe, D., et al.\ 2009, \apj, 690, 1313
\bibitem[Gezari et al.(2012)]{gez12} Gezari, S., Chornock, 
R., Rest, A., et al.\ 2012, \nat, 485, 217
\bibitem[Gezari et al.(2015)]{Gezari2015} Gezari, S., et al.\, in prep
\bibitem[Gezari et al.(2015a)]{Gezari2015a} Gezari, S., et al.~2015, \apj, 804, 28
\bibitem[Gezari et al.(2015b)]{Gezari2015b} Gezari, S., Chornock, R., Lawrence, A., Rest, A., Berger, E., Challis, P. M., \& Narayan, G.~2015, ApJ Letters, submitted
\bibitem[Greiner et al.(2008)]{gre08} Greiner, J., Bornemann, W., Clemens, C.,
et al.\ 2008, \pasp, 120, 405
\bibitem[Grupe et al.(1995)]{1995A&A...299L...5G} Grupe, D., Beuermann, K.,
Mannheim, K., et al.\ 1995, \aap, 299, L5 \bibitem[Grupe et al.(2015)]{gru15} Grupe, D., Komossa, S., 
\& Saxton, R.\ 2015, \apjl, 803, L28  
\bibitem[Houck et al.(2004)]{hou04} Houck, J.~R., Roellig, T.~L., van Cleve, J.,
et al.\ 2004, \apjs, 154, 18
\bibitem[Immler \& Lewin(2003)]{imm03} Immler, S., \& Lewin,
W.~H.~G.\ 2003, Supernovae and Gamma-Ray Bursters, 598, 91 
\bibitem[Immler 
\& Pooley(2007)]{imm07} Immler, S., \& Pooley, D.\ 2007, The Astronomer's
Telegram, 1004, 1 
\bibitem[Komossa \& Greiner(1999)]{kom99} Komossa, S., \& Greiner,
J.\ 1999, \aap, 349, L45 
\bibitem[Komossa et al.(2008)]{kom08} Komossa, S., Zhou, H., Wang, T., et al.\ 2008, \apjl, 678, L13 \bibitem[Komossa et al.(2009)]{kom09} Komossa, S., Zhou, H., Rau, A., et al.\ 2009, \apj, 701, 105 
\bibitem[Lang et al.(2010)]{lan10} Lang, D., Hogg, D.~W., Mierle, K., Blanton, M., \& Roweis, S.\ 2010, \aj, 139, 1782
\bibitem[Markwardt et al.(2002)]{mar02} Markwardt, C.~B., Swank,
J.~H., Strohmayer, T.~E., in 't Zand, J.~J.~M., \& Marshall, F.~E.\ 2002, \apjl, 575, L21 
\bibitem[Martin et al.(2005)]{mar05} Martin, D.~C., Fanson, 
J., Schiminovich, D., et al.\ 2005, \apjl, 619, L1
\bibitem[Nikutta et al.(2014)]{nik14} Nikutta, R., Hunt-Walker,
N., Nenkova, M., Ivezi{\'c}, {\v Z}., \& Elitzur, M.\ 2014, \mnras, 442, 3361
\bibitem[Obri{\'c} et al.(2006)]{obr06} Obri{\'c}, M., Ivezi{\'c},
{\v Z}., Best, P.~N., et al.\ 2006, \mnras, 370, 1677
\bibitem[Raskin et al.(2013)]{ras13} Raskin, G., Bloemen, S.,
Morren, J., et al.\ 2013, \aap, 559, A26
\bibitem[Rest et al.(2011)]{Rest2011}Rest, A., Foley, R.~J., Gezari, S., et al.\ 2011, \apj, 729, 88
\bibitem[Saxton et al.(2012)]{sax12} Saxton, R.~D., Read, A.~M., Esquej, P., et al.\ 2012, \aap,
541, AA106
\bibitem[Saxton et al.(2015)]{2015MNRAS.454.2798S} Saxton, R.~D., Motta, 
S.~E., Komossa, S., \& Read, A.~M.\ 2015, \mnras, 454, 2798  
\bibitem[Schlegel \& Petre(2006)]{sch06} Schlegel, E.~M., \&
Petre, R.\ 2006, \apj, 646, 378 
\bibitem[Sesar et al.(2011)]{ses11} Sesar, B. et al. 2011, \aj , 142, 190
\bibitem[Skrutskie et al.(2006)]{skr06} Skrutskie, M.~F., Cutri, R.~M.,
Stiening, R., et al.\ 2006, \aj, 131, 1163 
\bibitem[Smith et al.(2007)]{smi07} Smith, N., Li, W., Foley,
R.~J., et al.\ 2007, \apj, 666, 1116 
\bibitem[Smith et al.(2008)]{smi08} Smith, N., Chornock, C., Li, W., et
al.~2008, ApJ, 686, 467
\bibitem[Smith et al.(2009)]{smi09} Smith, N., Silverman, J.~M.,
Chornock, R., et al.\ 2009, \apj, 695, 1334
\bibitem[Stokes et al.(2000)]{sto00} Stokes, G.~H., Evans, J.~B., Viggh,
H.~E.~M., Shelly, F.~C.,\& Pearce, E.~C.\ 2000, \icarus, 148, 21 
\bibitem[Tody(1993)]{tod93} Tody, D.\ 1993, Astronomical Data 
Analysis Software and Systems II, 52, 173 
\bibitem[Turatto et al.(1993)]{Turatto1993} Turatto, M., Capellaro, E., Danziger, I. J., Benetti, S., Gouiffes, C., \& della Valle, M.~1993, \mnras, 262, 1993
\bibitem[van Velzen et al.(2011)]{vanVelzen2011} van Velzen, S., et al.~2011, \apj, 741, 73
\bibitem[Voges et al.(1999)]{vog99} Voges, W., Aschenbach, B., Boller, T., et
al.\ 1999, \aap, 349, 389 
\bibitem[Yang et al.(2013)]{yan13} Yang, C.-W., Wang, T.-G., Ferland, G., et
al.\ 2013, \apj, 774, 46
\bibitem[Wang et al.(2011)]{Wang2011} Wang, T. G., Zhou, H. Y., Wang, L., Lu, H. L., \& Xu, D. W.~2011, \apj, 740, 85
\bibitem[Wang et al.(2012)]{Wang2012} Wang, T. G., Zhou, H. Y., Komossa, S.,
Wang, H. Y., Yuan, W., \& Yang, C.~2012, \apj, 749, 115
\bibitem[Wright et al.(2010)]{2010AJ....140.1868W} Wright, E.~L., 
Eisenhardt, P.~R.~M., Mainzer, A.~K., et al.\ 2010, \aj, 140, 1868 

\end{thebibliography}
\end{document}